\documentclass[aps,prl,preprint,groupaddress]{revtex4}

\newcommand {\EMB}{EuMnBi$_2$ }

\usepackage{graphicx}
\usepackage{bm}
\usepackage{pifont}
\usepackage{amssymb, amsmath}
\usepackage{dcolumn}
\usepackage{color}

\begin{document}
\title{
Impact of antiferromagnetic order on Landau level splitting of quasi-two-dimensional Dirac fermions in EuMnBi$_2$
}
\author{H. Masuda$^1$}
\author{H. Sakai$^{2,3}$}\email[Corresponding author: ]{sakai@phys.sci.osaka-u.ac.jp}
\author{M. Tokunaga$^4$}
\author{M. Ochi$^{2}$}
\author{H. Takahashi$^1$}
\author{K. Akiba$^4$}
\author{A. Miyake$^4$}
\author{K. Kuroki$^{2}$}
\author{Y. Tokura$^{1,5}$}
\author{S. Ishiwata$^{1,3}$}
\affiliation{$^1$Department of Applied Physics, University of Tokyo, Tokyo 113-8656, Japan.\\
$^2$Department of Physics, Osaka University, Toyonaka, Osaka 560-0043, Japan.\\
$^3$PRESTO, Japan Science and Technology Agency, Kawaguchi, Saitama 332-0012, Japan.\\
$^4$The Institute for Solid State Physics, University of Tokyo, Kashiwa 277-8581, Japan.\\
$^5$RIKEN Center for Emergent Matter Science (CEMS), Wako 351-0198, Japan.}
\begin{abstract}
We report spin-split Landau levels of quasi-two-dimensional Dirac fermions in a layered antiferromagnet EuMnBi$_2$, as revealed by interlayer resistivity measurements in a tilted magnetic field up to $\sim$35 T.
The amplitude of Shubnikov-de Haas (SdH) oscillation in interlayer resistivity is strongly modulated by changing the tilt angle of the field, i.e., the Zeeman-to-cyclotron energy ratio.
The effective $g$ factor estimated from the tilt angle, where the SdH oscillation exhibits a phase inversion, differs by approximately 50\% between two antiferromagnetic phases.
This observation signifies a marked impact of the magnetic order of Eu sublattice on the Dirac-like band structure.
The origin may be sought in strong exchange coupling with the local Eu moments, as verified by the first-principles calculation. 
\end{abstract}
\maketitle
%
Dirac fermions in solids have attracted extensive attention for their unusual quantum transport phenomena\cite{Vafek2014AnnuRev}, typified by a half-integer quantum Hall effect in graphene\cite{Novoselov2005Nature,Zhang2005Nature}.
As a bulk analogue of graphene, so-called Dirac and Weyl semimetals hosting linear energy dispersion are recently of particular interest\cite{Armitage2018RMP}.
One of the greatest advantages of their bulk form is the interplay of relativistic quasiparticles with magnetism, which potentially leads to novel (spin)electronic applications\cite{Smejkal2017PSSR,Smejkal2018NatPhys}.
Recently, a few candidates of Dirac or Weyl magnets were reported, as exemplified by Mn$_3$Sn\cite{Nakatsuji2015Nature,Nayak2016SciAdv}, GdPtBi\cite{Hirschberger2016NatMat,Suzuki2016NatPhys}, and pyrochlore iridates\cite{Wan2011PRB,Krempa2014ARCMP}.
Some of these materials were found to exhibit peculiar magneto-transport phenomena, such as large anomalous Hall effects\cite{Nakatsuji2015Nature,Nayak2016SciAdv,Suzuki2016NatPhys} and chiral anomalies\cite{Hirschberger2016NatMat,Kuroda2017NatMat}, consistent with the theoretically-predicted Weyl semimetal states.
For exploring their potential applications, the roles of magnetic order on the topological electronic and transport properties need to be experimentally elucidated, which remains a work in progress\cite{Kuroda2017NatMat}.
%
\par
\begin{figure}
\begin{center}
\includegraphics[width=.9\linewidth]{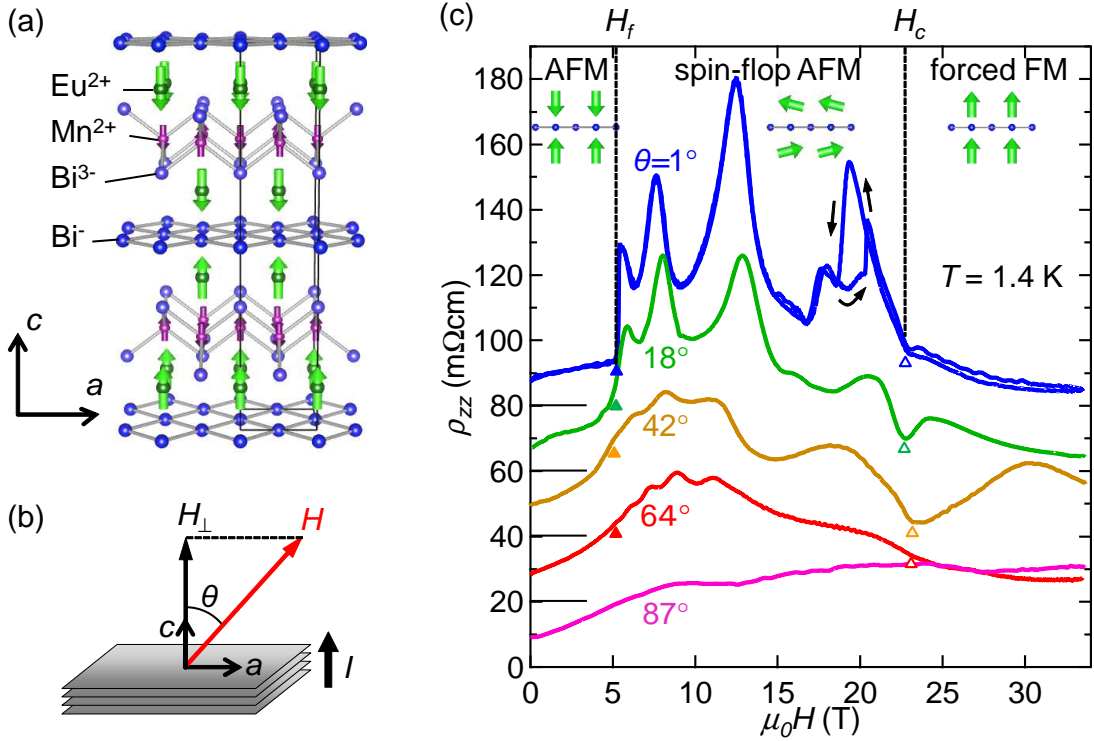}
\caption{
(Color online)
(a) Schematic illustration of the crystal and magnetic structure at 0 T for \EMB\cite{Masuda_EMB,SMB_neutron}. (b) Geometry of the interlayer transport measurement in a tilted magnetic field in the $a$-$c$ plane, where $\theta$ is an angle between the field and the $c$ axis. (c) Field profile of $\rho_{zz}$ at 1.4 K for selected values of $\theta$. For clarity, the curves are shifted vertically by 20 m$\Omega$cm. For $\theta\!<\! 64^{\circ}$, the closed triangle denotes $H_{\rm f}$ while the open triangle denotes $H_{\rm c}$. The positions of $H_{\rm f}$ and $H_{\rm c}$ are determined as the fields where $\rho_{zz}$ shows a jump and drop in the field-increasing run, respectively. For details, see supplementary Fig. S1. The magnetic order of the Eu sublattice for each antiferromagnetic phase is shown schematically in the inset.
}
\label{fig:intro}
\end{center}
\end{figure}

%
$A$Mn$X_2$ ($A$: alkaline-earth and rare-earth ions, $X$: Bi and Sb)\cite{Park2011PRL,Wang2011PRB,WangPetrovic2011PRB,JiaPRB2014,Li2016PRB,Wang2016CPB,May_EMB,Masuda_EMB,Borisenko2015arxiv,Wang2016PRB,Liu2017NatMater,Liu2015SR,Huang2016PNAS} is also promising as a fertile ground for magnetic Dirac materials, since the crystal structure consists of an alternate stack of a two-dimensional (2D) Dirac fermion conduction layer ($X^{-}$ square net)\cite{Lee2013PRB,Farhan2014JPC} and a magnetic insulating layer ($A^{2+}$-Mn$^{2+}$-$X^{3-}$) [see Fig. \ref{fig:intro}(a)].
Among them, EuMnBi$_2$ is a rare compound that exhibits quantum transport of Dirac fermions coupled with the field-tunable magnetic order.
In this compound, the interlayer coupling between each Dirac fermion (Bi) layer is dramatically suppressed by the flop of the antiferromagnetically-ordered Eu moments [Fig. \ref{fig:intro}(a)].
The enhanced two dimensionality leads to the giant magnetoresistance effects\cite{May_EMB,Masuda_EMB} and the quantum oscillation phenomena\cite{Masuda_EMB} that strongly depend on the magnetic order of the Eu sublattice.
However, in spite of such a marked impact of magnetism on the transport properties, it remains elusive how and to what extent the Dirac-like band dispersion is affected.
%
\par
%
To reveal the coupling between the band structure and magnetic order, the Landau level quantization in a magnetic field can be a powerful probe, since it exhibits the energy splitting due to Zeeman and exchange coupling as well as electron-electron interaction.
As demonstrated in the conventional 2D electron gas in semiconductor heterostructures\cite{Ando1982RMP} and semimagnetic quantum wells\cite{Teran2002PRL}, the detailed analyses on the splitting provide lots of information on the band parameters and magnetism of the system, which have been recently performed for graphene\cite{Zhang2006PRL,Young2012NatPhys} and several Dirac semimetals\cite{Jeon2014NatMater,Xiang2015PRL,Liu2016NatCom}.
Also for EuMnBi$_2$, clear Landau level splitting was observed in the SdH oscillation in resistivity\cite{Masuda_EMB}, the origin of which has not been clarified so far.
In this Letter, we clarify that the Landau level splitting in EuMnBi$_2$ is primarily of spin origin, on the basis of the systematic measurements of the SdH oscillations in tilted magnetic fields.
The field-angle dependence of SdH oscillations have revealed the effective $g$ factors for the Dirac fermions, which strongly depends on the antiferromagnetic (AFM) order of the Eu sublattice.
As a plausible explanation, we discuss the exchange coupling between Dirac fermions and local Eu moments by considering the results of the first-principles calculations.
%
%
\par
%
For investigating the fine structures of Landau levels, we have here adopted the measurements of interlayer resistivity $\rho_{zz}$.
This is because the high-resistive $\rho_{zz}$ has a much better S/N ratio than that achieved in the in-plane resistivity $\rho_{xx}$.
A rotation of magnetic field is also important in the present study.
In 2D systems, the ratio of the cyclotron energy $E_{\rm c}$ to the Zeeman energy $E_{\rm Z}$ can be tuned by changing the tilt angle of the field from the normal to the 2D plane ($\theta$); $E_{\rm c}$ is proportional to $H_\perp=H\cos\theta$ [the field component perpendicular to the 2D plane, see Fig. \ref{fig:intro}(b)], while $E_{\rm Z}$ is proportional to $H$ (the total field).
The combination of these techniques allow us to elucidate the mechanism of the Landau level splitting and hence the microscopic nature of the Dirac fermions in EuMnBi$_2$, as described below.
%
\par
%
Figure \ref{fig:intro}(c) shows the field dependence of interlayer resistivity $\rho_{zz}$ for EuMnBi$_2$ up to 35 T at selected tilt angles of the field.
We first review the transport features for the field parallel to the $c$ axis (at $\theta=1^\circ$).
With increasing the field, $\rho_{zz}$ exhibits a steep jump at the spin-flop transition of the Eu sublattice ($H_f\sim 5.3$ T), followed by large SdH oscillations.
In the forced ferromagnetic (FM) phase above $H_{\rm c}\!\sim\! 22$ T, however, the value of $\rho_{zz}$ significantly decreases, indicating that $\rho_{zz}$ is specifically enhanced in the spin-flop AFM phase.
There, the Dirac fermions in the Bi layer are strongly confined in two dimension, resulting in the signature of multilayer half-integer quantum Hall effect in the in-plane conductions\cite{Masuda_EMB}.
%
\par
%
Similar enhancement in $\rho_{zz}$ in the spin-flop AFM phase was observed at $\theta$ up to $\sim$65$^\circ$, which is gradually reduced with increasing $\theta$.
Concomitantly, the spin-flop transition at $H_{\rm f}$ is less sharp at high $\theta$, which is still discernible up to $\theta\!=\!64^{\circ}$ as denoted by closed triangles in Fig. \ref{fig:intro}(c) (for the determination of $H_{\rm f}$, see supplementary Fig. S1).
The manner of the SdH oscillation is also strongly dependent on $\theta$, whereas the values of $H_{\rm f}$ and $H_{\rm c}$ are almost independent of $\theta$.
Note here that, in addition to the SdH oscillation, a hysteretic resistivity anomaly is discernible around 20 T at $\theta$=1$^{\circ}$, which immediately disappears when $\theta$ increases up to 18$^{\circ}$.
At present, the origin of this highly-$\theta$-sensitive anomaly remains unclear, the study of which is beyond the scope of this paper.
In the following, we shall focus on the $\theta$ dependence of the SdH oscillations in $\rho_{zz}$.
%
\begin{figure}
\begin{center}
\includegraphics[width=.65\linewidth]{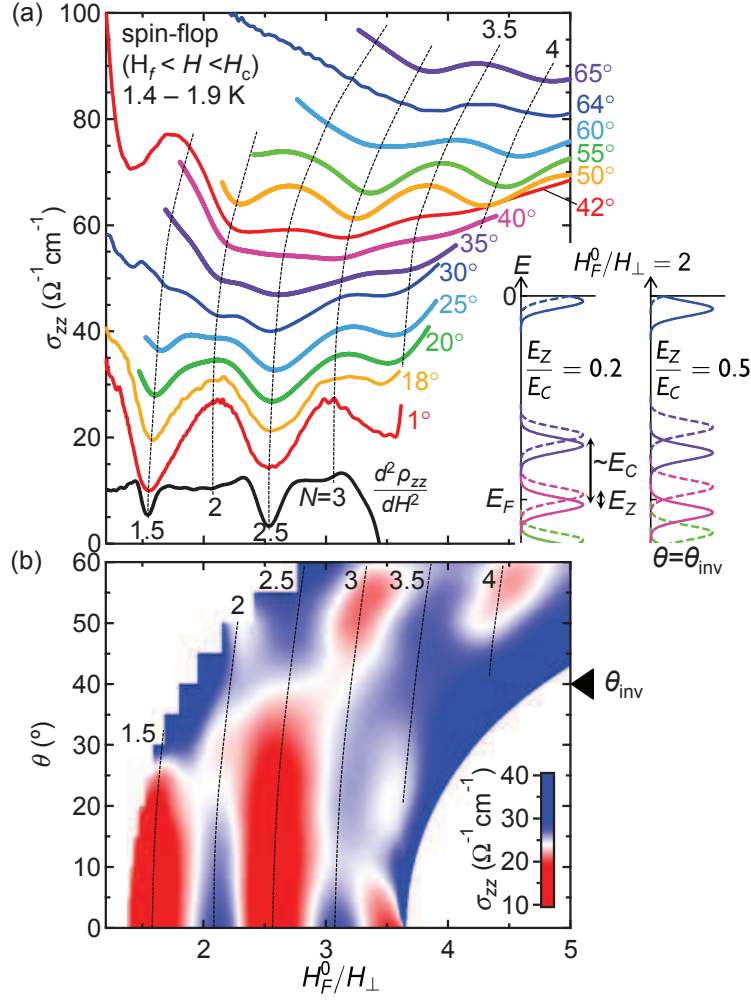}
\caption{
(Color online)
(a) $\sigma_{zz}$ versus $H_F^0/H_\perp$ at $\theta$=1$^{\circ}$-65$^{\circ}$ in the spin-flop AFM phase ($H_f<H<H_c$), where $H_F^0$ denotes the SdH frequency for the field parallel to the $c$ axis. The curves at $\theta\ge 18^{\circ}$ are shifted upward for clarity. At the bottom of the panel, the second field derivative $d^2\rho_{zz}/dH^2$ at $\theta\!=\! 1^\circ$ is shown. Vertical dotted lines are guides to the eye showing the positions of the maxima and minima of the SdH oscillation, where $N$ denotes the Landau index.
Inset: Schematic of the density of states for spin-split Landau levels for a 2D massless Dirac fermion as a function of energy $E$ for $H_F^0/H_\perp=2$, where $E_{\rm F}$ is set negative corresponding to the hole carrier system.
The value of $E_{\rm Z}/E_{\rm c}$ can be tuned by tilting the field, where $E_{\rm Z}\!=\!g^{\ast}\mu_{\rm B}B$ is the Zeeman energy, $E_{\rm c}\!\equiv\! e\hbar B_\perp /m_{\rm c}$ the effective cyclotron energy, and $m_{\rm c}$ the cyclotron mass $m_{\rm c}\!=\! E_{\rm F}/v_{\rm F}^2$. $E_{\rm Z}/E_{\rm c}\!=\! 0.2$ (left) and 0.5 (right).
For details of the calculation, see the main text and supplementary Fig. S4.
(b) Color plot of $\sigma_{zz}$ as functions of $H_F^0/H_\perp$ and $\theta$. 
$\theta_{\rm inv}$ indicated by the triangle corresponds to $\theta$ where the phase of the SdH oscillation is inverted and $E_{\rm Z}/E_{\rm c}$ is nearly 0.5.
}
\label{fig:AMR_SF}
\end{center}
\end{figure}
%
\par
%
We first show in Fig. \ref{fig:AMR_SF}(a) the features of the Landau levels in the spin-flop AFM phase ($H_f<H<H_c$) by presenting the $\theta$ dependence of interlayer conductivity $\sigma_{zz}=1/\rho_{zz}$\cite{note}.
The horizontal axis of Fig. \ref{fig:AMR_SF}(a) denotes $H_F^0/H_\perp$, the normalized filling factor for a 2D system\cite{Lukyanchuk2006PRL,Masuda_EMB}, where $H_F^0$(=19.3 T) is the SdH frequency for the field parallel to the $c$ axis (Fig. S3)\cite{SM}.
At $\theta$=1$^{\circ}$, $\sigma_{zz}$ shows the minima at $H_F^0/H_\perp\simeq 1.5, 2.5, 3.5$, which coincides with the oscillations in $\sigma_{xx}$ and $\rho_{xx}$\cite{Masuda_EMB}.
Since the deep minima in $\sigma_{zz}$ and $\sigma_{xx}$ indicate the quantum Hall states\cite{Druis1998PRL,Kuraguchi2000PhysicaE,Kawamura2001PhysicaB}, the corresponding $H_F^0/H_\perp$ should be given by $H_F^0/H_\perp\!=\! N + 1/2 - \gamma$, where $N$ is the Landau index and $\gamma$ is the phase factor expressed as $\gamma\!=\! 1/2 - \phi_B/2\pi$ with $\phi_B$ the Berry's phase\cite{Mikitik1999PRL}.
The $\sigma_{zz}$ minima occurring at half-integer multiples of $H_F^0/H_\perp$ thus lead to $\gamma\!\sim\!0$, i.e., the nontrivial $\pi$ Berry's phase in EuMnBi$_2$.
In multilayer quantum Hall systems, it was reported that a chiral surface state contributes partly to the interlayer transport in the quantum Hall states (i.e., $\sigma_{zz}$ minima)\cite{Druis1998PRL,Kuraguchi2000PhysicaE,Kawamura2001PhysicaB}, which does not affect the frequency or phase of the SdH oscillation discussed below.
When $\theta$ increases, the frequency of the SdH oscillation increases in proportion to $1/\cos\theta$ [Fig. S3(c)]\cite{SM}, indicating highly 2D nature of the Fermi surface.
This results in the almost $\theta$-independent oscillation period when plotted as a function of $H_F^0/H_\perp$, as highlighted by the vertical dotted lines up to $\theta\sim 50^{\circ}$ in Fig. \ref{fig:AMR_SF}(a).
For $\theta\!\ge\! 55^{\circ}$, however, the frequency gradually deviates from the $1/\cos\theta$ scaling presumably due to a weak warping of the Fermi surface caused by the non-zero interlayer coupling.
%
\par
%
The most salient feature of the SdH oscillation is that the amplitude significantly varies with $\theta$.
With increasing $\theta$ up to 35$^\circ$-40$^\circ$, the amplitude progressively decreases to nearly zero.
Above $\theta\!=\! 40^\circ$, the amplitude again increases but with an inverted phase.
The observed $\theta$ dependence of the SdH amplitude is well explained by considering the spin splitting of the Landau levels due to $E_{\rm Z}$ as follows\cite{Si100_spinsucceptibility_Fang1968, AlAs_spinsucceptibility_Vakili2004, ZnO_spinsucceptibility_Tsukazaki2008}.
When $E_{\rm Z}/E_{\rm c}$ is smaller than unity [e.g., $E_{\rm Z}/E_{\rm c}\!=\! 0.2$ in the inset (left) to Fig. \ref{fig:AMR_SF}(a)], the Landau level exhibits a weak spin splitting, which is barely discernible at $\theta\!\sim\! 1^\circ$ when plotted in the form of $d^2\rho_{zz}/dH^2$ [Fig. \ref{fig:AMR_SF}(a)]\cite{Masuda_EMB} .
With increasing $E_{\rm Z}/E_{\rm c}$ by tilting the field, the magnitude of the spin splitting increases, resulting in the reduction in amplitude of the SdH oscillation.
Around $\theta\!=\! 40^\circ$, the amplitude reaches the minimum, which corresponds to $E_{\rm Z}/E_{\rm c}\!=\! 0.5$ [the inset (right) to  Fig. \ref{fig:AMR_SF}(a)].
A further increase in $E_{\rm Z}/E_{\rm c}$ leads to crossing of the neighboring Landau levels with opposite spins, which results in the enhanced SdH oscillation with an inverted phase, as observed at $\theta\!>\! 50^\circ$.
Note here, since the energy spacing of Landau levels for a 2D Dirac fermion is not uniform (i.e., $E_{\rm c}$ is dependent on $N$), we need to effectively define $E_{\rm c}\!\equiv\! e\hbar B_\perp /m_{\rm c}$ by using a semiclassical expression of the cyclotron mass $m_{\rm c}\!=\! E_{\rm F}/v_{\rm F}^2$ with $v_{\rm F}$ and $E_{\rm F}$ being the Fermi velocity and Fermi energy, respectively\cite{Novoselov2005Nature,Zhang2005Nature}.
In this scheme, the Landau level crossing in the SdH oscillation occurs when $E_{\rm Z}/E_{\rm c}\!=\! 1$ irrespective of $N$, as in the case for a normal 2D electron gas [for details, see Fig. S4(b)]\cite{SM}.
%
\par
%
To highlight the $\theta$ dependence of the SdH oscillations, we present a contour plot of $\sigma_{zz}$ as functions of $H_F^0/H_\perp$ and $\theta$ in Fig. \ref{fig:AMR_SF}(b).
It is clear that the phase of the SdH oscillation is inverted around $\theta_{\rm inv}\!\sim\! 40^{\circ}$, accompanied by the minimum amplitude.
As shown in supplementary Fig. S4(c)\cite{SM}, this plot is nicely reproduced by calculating the density of states of spin-split Landau levels, where $E_{\rm Z}/E_{\rm c}\!=\! 0.5$ corresponds to $\theta\!=\! \theta_{\rm inv}$\cite{note2}.
Noting that $E_{\rm Z}/E_{\rm c}\!=\! g^\ast m_{\rm c}/2m_0\cos\theta$, this relation gives $\cos\theta_{\rm inv}\!=\! g^\ast m_{\rm c}/m_0$, where $g^\ast$ is the effective $g$ factor and $m_0$ is the bare electron mass.
By substituting the experimental value ($\theta_{\rm inv}\!=\! 40^\circ\!\pm\! 5^\circ$), we obtain $g^\ast m_{\rm c}/m_0\!=\! 0.77(6)$.
The value of $m_{\rm c}/m_0$ is independently estimated to be 0.122(2) from the temperature dependence of the SdH oscillations at $\theta\!=\! 0^\circ$ based on the standard Lifshitz-Kosevich formula (Fig. S8)\cite{SM}, which results in $g^\ast\!=\! 6.6(6)$.
The obtained $g^\ast$ is much larger than 2, reflecting strong spin-orbit coupling inherent to Bi atom.
Additionally, it is presumable that the exchange interaction with the local Eu moments plays a significant role, since net magnetization is non-zero in the spin-flop AFM phase, as discussed later.
%
\par
%
\begin{figure}
\begin{center}
\includegraphics[width=.6\linewidth]{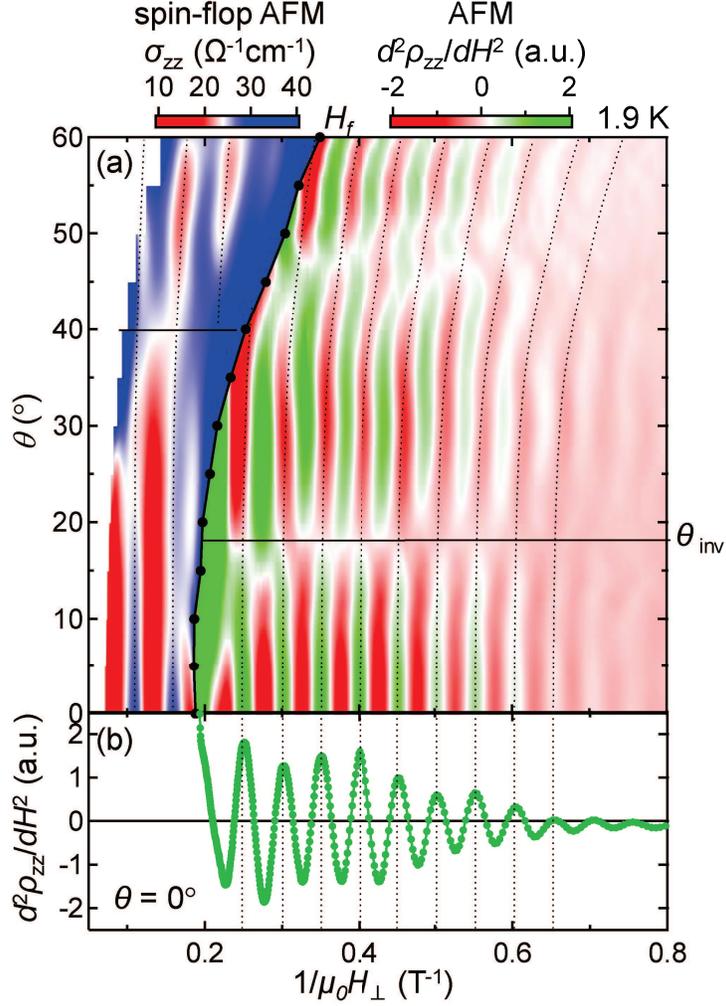}
\caption{
(Color online)
(a) Color plot of $d^2\rho_{zz}/dH^2$ as functions of $H_F^0/H_\perp$ and $\theta$ in the AFM phase (for $H\!<\! H_f$). To compare the $\theta$ dependence, the $\sigma_{zz}$ data in the spin-flop AFM phase (for $H\!>\! H_f$) are plotted together.
The position of $H_{\rm f}$ at each $\theta$ is denoted by a closed circle, which is determined as the field where $\rho_{zz}$ shows a steep increase (see supplementary Fig. S1).
The horizontal line denotes $\theta_{\rm inv}$ for each phase.
(b) Profile of $d^2\rho_{zz}/dH^2$ versus $H_F^0/H_\perp$ for $\theta\!=\!0^\circ$ ($H\!<\! H_f$).}
\label{fig:AMR_AFM}
\end{center}
\end{figure}
%
Next, we shall show the Landau level splitting in the AFM phase ($H\!<\! H_f$), where the amplitude of SdH oscillation is largely reduced as compared with the spin-flop AFM phase.
Nonetheless, the oscillation is discernible above $\sim$1.3 T, as shown in Fig. \ref{fig:AMR_AFM}(b) where $d^2\rho_{zz}/dH^2$ is plotted for clarity.
The weakly beating amplitude presumably signifies the superposition of maximum and minimum cyclotron orbits arising from a slightly warped cylindrical Fermi surface.
To summarize the $\theta$ dependence of SdH oscillation, we show in Fig.\ref{fig:AMR_AFM}(a) the color contour plots of $d^2\rho_{zz}/dH^2$ and $\sigma_{zz}$ for the AFM and spin-flop AFM phases, respectively, as functions of $H_F^0/H_\perp$ and $\theta$.
The SdH oscillation in the AFM phase has several common features with that in the spin-flop AFM phase; the period of the SdH oscillation is nearly independent of $\theta$ when plotted versus $1/H_\perp$, reflecting a quasi-2D Fermi surface.
In addition, the spin splitting of the Landau levels makes the oscillation amplitude dependent on $\theta$, leading to the phase inversion at $\theta_{\rm inv}$ (a horizontal line).
However, the value of $\theta_{\rm inv}$ is substantially different in the two phases; $\theta_{\rm inv}\!\sim\! 18^\circ$ for the AFM phase while $\theta_{\rm inv}\!\sim\! 40^\circ$ for the spin-flop AFM phase.
This results in $g^\ast m_c/m_0=\cos\theta_{\rm inv}=0.95(1)$ for the AFM phase ($\theta_{\rm inv}\!=\! 18^\circ \pm 2^\circ)$, which is approximately 25\% larger than that for the spin-flop AFM phase.
%
\par
%
In Table \ref{tb:list}, we compare the band parameters estimated from the SdH oscillation for each AFM phase.
The cross section of quasi-2D Fermi surface $S_{\rm F}$ deduced from the SdH frequency ($H_F^0$) is almost the same for both AFM phases, whereas the values of $m_{\rm c}$ and $g^\ast$ significantly depend on the AFM order.
Since the AFM phase hosts larger $g^\ast m_{\rm c}/m_0$ and smaller $m_{\rm c}/m_0$ than the spin-flop AFM phase, the resultant $g^\ast$ value for the former phase is approximately 50\% larger than that for the latter phase.
These facts indicate that the Dirac-like band for EuMnBi$_2$ is largely modulated by the AFM order of Eu sublattice.
%
\begin{table}[t]
  \begin{center}
  \caption{Experimentally determined band parameters for the AFM and spin-flop AFM phases. For the estimation of $S_F$ and $m_c$, see supplementary Figs. S3 and S6$-$S10.}\label{tb:list}
  \begin{tabular}{| c || c | c | c | c |} \hline
                    &\hfil $S_F\ {\rm (nm^{-2})}$  \hfil&\hfil $g^\ast m_c/m_0$ \hfil&\hfil $m_c/m_0$   \hfil&\hfil $g^\ast$ \hfil\\\hline\hline
     AFM        &\hfil 0.186                              \hfil&\hfil 0.95(1)          \hfil&\hfil 0.097(2) \hfil&\hfil 9.8(4)     \hfil\\
\hline
     spin-flop  &\hfil 0.191                              \hfil&\hfil 0.77(1)          \hfil&\hfil 0.122(2) \hfil&\hfil 6.6(4)     \hfil\\\hline
  \end{tabular}
  \end{center}
\end{table}
%
\begin{figure}
\begin{center}
\includegraphics[width=.9\linewidth]{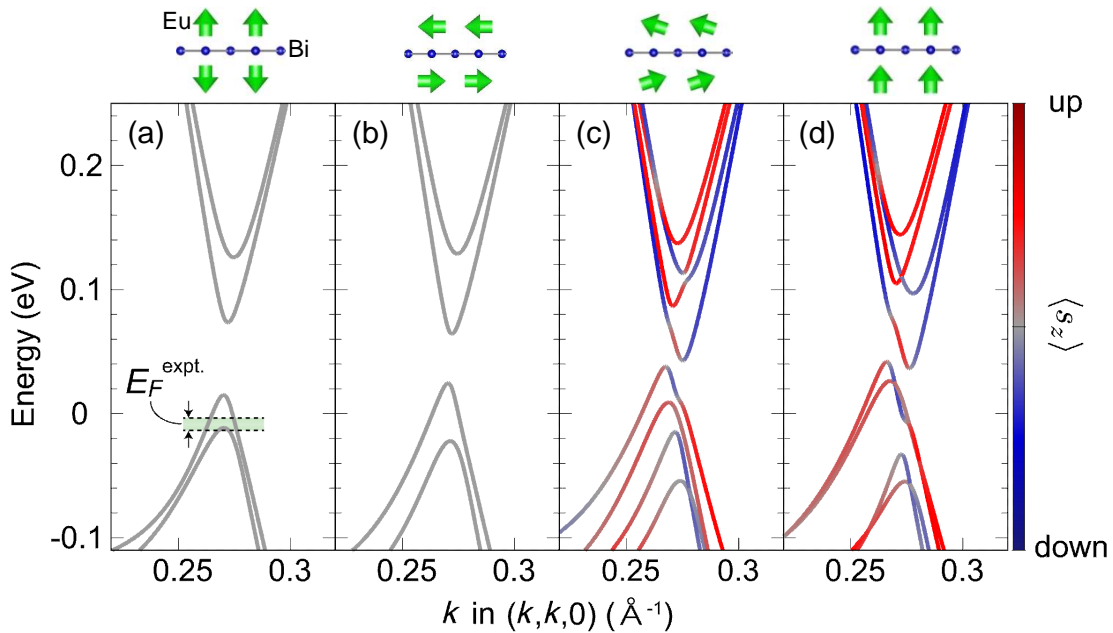}
\caption{
(Color online)
Calculated Dirac-like band structures along the $\Gamma$-M line for various magnetic states in EuMnBi$_2$. (a) AFM, (b, c) spin-flop AFM, and (d) forced FM states. In (b), the Eu moment is along the $a$ axis, while in (c) it is inclined at an angle of $\sim$47$^\circ$ to the $c$ axis on the $ac$ plane. The spin polarization $\langle s_z\rangle$ of each band is represented by red (up) and blue (down) colors. Schematic illustration of the Eu moments adjacent to the Bi layer is also shown. The Fermi energy $E_{\rm F}$ estimated from the experimental SdH oscillation is denoted by the shaded area in (a). For details, see supplementary Fig. S12}
\label{fig:calculation}
\end{center}
\end{figure}
%
\par
%
First-principles calculations indeed reproduce such a marked dependence of the band structure on the magnetic state, as shown in Fig. \ref{fig:calculation}, where the Dirac-like bands near $E_{\rm F}$ are displayed\cite{note5}.
Note that two sets of bands arise from the unit cell doubling along the $c$ axis to represent the AFM order of Eu moments, which is adopted to the forced FM state in common\cite{SM}.
In addition to a small change upon the spin flop of the Eu moments [Figs. \ref{fig:calculation}(a) and (b)], the splitting of red-colored (spin up) and blue-colored (spin down) bands progressively evolves, as the net magnetization (i.e., the canting of the Eu moment) increases in the spin-flop AFM phase [Figs. \ref{fig:calculation}(b)$-$(d)].
Since the present calculation does not take $E_{\rm Z}$ into consideration, this large spin splitting originates from the exchange coupling of the Dirac fermion with the local Eu moments ($E_{\rm ex}$), which can be expressed as $E_{\rm ex}\!=\! J\langle S\rangle \!=\! J\chi H/g_{J}$, where $J$ is the exchange integral, $\langle S\rangle$ is the component of Eu$^{2+}$ spin along the field, $g_{J}(=2)$ is the Land\'e $g$ factor for Eu$^{2+}$, and $\chi$ is the magnetic susceptibility.
In the AFM phase, since the field is parallel to the easy axis of Eu spins, $\chi$ is a small parallel susceptibility and hence $E_{\rm ex}$ is negligible.
On the other hand, in the spin-flop AFM phase, where the Eu spin axis changes to be transverse to the field, $\chi$ corresponds to a much larger transverse susceptibility\cite{note3}. 
In the latter phase, the Landau level splitting is caused by $E_{\rm ex}$ as well as $E_{\rm Z}$, which renormalizes the $g^\ast$ value.
From the energy splitting shown in Fig. \ref{fig:calculation}(d), we obtain $E_{\rm ex}\! =\!$ 50$-$80 meV\cite{note4} for $\langle S\rangle$=7/2 (i.e., $J\! =\! 14$$-$23 meV), which is comparable to (or even larger than) $E_{\rm Z}\!\sim\! 13$ meV at $H\!=\! H_{\rm c}\!\sim\! 22$ T for $g^\ast\!\sim\! 10$.
Thus, the exchange coupling should appreciably contribute to the observed apparent change in $g^\ast$ upon the AFM phase.
The reduction of $g^\ast$ in the spin-flop AFM phase implies that the sign of $J$ is opposite to that of pristine $g^\ast$, although a more quantitative estimation of these parameters is a future subject.
%
\par
%
In conclusion, we observed spin-split Landau levels of quasi-two-dimensional Dirac fermions in a bulk antiferromagnet EuMnBi$_2$, which markedly depend on the field-tunable magnetic order of Eu moments.
In addition to Zeeman splitting relevant to the large $g$ factor, the Dirac-like band exhibits substantial exchange splitting due to the coupling with the local Eu moments.
Such an interplay of the spin-orbit and exchange interactions in the present compound yields novel correlated Dirac fermion states in a solid, offering a promising approach to emerging topological spintronics.
%
\begin{acknowledgments}
The authors thank T. Osada, A. Tsukazaki and Y. Fuseya for helpful discussions.
This work was partly supported by PRESTO, JST (Nos. JPMJPR16R2 and JPMJPR1412), Grant-in-Aid for Young Scientists A (No. 16H06015), Grant-in-Aid for Scientific Research A (No. 17H01195), and the Asahi Glass Foundation.
\end{acknowledgments}
%


\begin{thebibliography}{}
\bibitem{note3} From the magnetization data\cite{Masuda_EMB}, we have estimated $\chi_{||}$ and $\chi_{\perp}$ to be $5.4\!\times\! 10^{-3}$ and $1.4\!\times\! 10^{-1}$ emu/mol at 2 K, respectively.
\end{thebibliography}
\end{document}